\documentclass[pra,aps,twocolumn,superscriptaddress,showpacs,tightenlines,amsmath]{revtex4}
\usepackage{epsfig}
\usepackage{amsmath,amssymb}

\begin{document}

\title{Factorization of unitary matrices induced by 3D anisotropic Ising interaction}

\author{Francisco Delgado}
\email{fdelgado@itesm.mx}
\affiliation{Escuela Nacional de Posgrado en Ciencias e Ingenier\'ia, Tecnol\'ogico de Monterrey, M\'exico.}
\affiliation{Departamento de F\'isica y Matem\'aticas, Tecnol\'ogico de Monterrey, Campus Estado de M\'exico, Atizap\'an, Estado de M\'exico, CP. 52926, M\'exico.}

\date{\today}

\begin{abstract}

Quantum computation is a continuously growing research area which is based on nature and resources of quantum mechanics, as superposition and entanglement. In its quantum circuits version, the use of convenient and appropriate gates is essential. But while those gates adopt  convenient forms for computational algorithms, their design depends on specific quantum systems and stuff being used. These gates need manage quantum systems based on physical interactions ruled by quantum Hamiltonians. With this, gates design is restricted to properties and limitations of interactions and physical elements being involved. This work shows how anisotropic Ising interactions, written in a non local basis, lets reproduce elementary operations in terms of which unitary processes can be factorized. In this sense, gates could be written as a sequence of pulses ruled by that interaction driven by magnetic fields, stating alternative results in quantum gates design for magnetic systems.



\end{abstract} 

\maketitle

\section{Introduction}
Quantum computation is the up most application goal of quantum mechanics. While this goal could be reached, in parallel, other useful appliances based on quantum information are being effectively developed and obtained, as cryptography and teleportation. Quantum gate array computation is the most common and clear approach in terms of proximity with classical computer science. This proximity has made that some concepts and specially some gates had been replicated from classical programming. Nevertheless, this gates should be reproduced by several physical interactions and resources as ion traps and electromagnetic cavities \cite{cirac1,turchette1}, Josephson junctions \cite{shnirman1}, nuclear magnetic resonance \cite{chuang1} and spins \cite{kane1,loss1,vrijen1}, thus as its translation is not always immediate, requiring complex control or iterative procedures.

Ising model \cite{ising1,brush1,baxter1,nielsen2} can be used as a simple approach between interaction of magnetic quantum objects (electronic gases, quantum dots, ions, etc.). Nielsen \cite{nielsen2} was the first reporting studies of entanglement in magnetic systems based on a two spin systems driven with an external magnetic field. One of the properties of this model is that it generates entanglement, which is one of the interesting properties of quantum mechanics noted since early times \cite{vneumann1, schrod1, schrod2, einstein1, schrod3}. This property is a central aspect in the most of quantum applications improving capacity and speed information processing \cite{jozsa1, jozsa2, bennet1}. In this sense, entanglement is an important aspect to codify and manage information with alternative methods to those of classical computer science, since fundamental proposals in Quantum Computation \cite{feyn1, deutsch1, steane1}, Quantum Cryptography \cite{bennet2, ekert1}, superdense coding \cite{bennet3} and teleportation \cite{bennet4}. Control of entanglement in its several variants can be achievable in Ising model through of driven magnetic fields being introduced on the physical system. Nevertheless some complex studies about this interaction in multipartite cases (some of them numerical more than analytical because of its complexity when number of parts grown), is still useful comprehend how this useful model could be used to reproduce efficient procedures in a couple of qubits at time to reproduce some basic elements in quantum gate array computation. Different models of Ising interaction (XX, XY, XYZ depending on interest of each author and physical systems being considered) are used to reproduce effects related with bipartite or multipartite systems \cite{berman1,wang2,aless1} and quantum dots \cite{brunner1, gau1}). 
  
Nowadays, quantum gate array computation is being experimentally explored in terms to adapt it to stuff in which it can be settled in terms particularly of noise control and reproduction of basic gates. It means, interactions able to be considered to reproduce it \cite{cirac1,turchette1,shnirman1,chuang1,kane1,loss1,vrijen1}. Some examples are quantum dots or electronic gases, which are developments towards a scalable spin-based quantum computer which can be controlled with electromagnetic interactions between neighboring spins, being believed able to obtain universal quantum operations \cite{recher1, saraga1, kopp1} in terms of DiVincenzo criteria \cite{vincenzo1} about reliability of state preparation, identification of well identified qubits and accurate quantum gate operations. The aim of this paper is to analyze how a set of anisotropic Ising model interactions recently reported \cite{delgadoA} can be applied to reproduce gates in terms of algebraic factorization driven by inhomogeneous magnetic fields. This factorization could be useful in quantum simulation to reproduce the evolution of a desired computational problem in order to solve it. One central aspect of this implementation is that analysis of dynamics is constructed on the non-local basis of classical Bell states which becomes outstanding by its regular algebraic structure, which fits to desired factorization presented here.

\section{Analytic evolution for anisotropic Ising model in three dimensions}

\begin{figure}[pb] 
\centerline{\psfig{width=2.8in,file=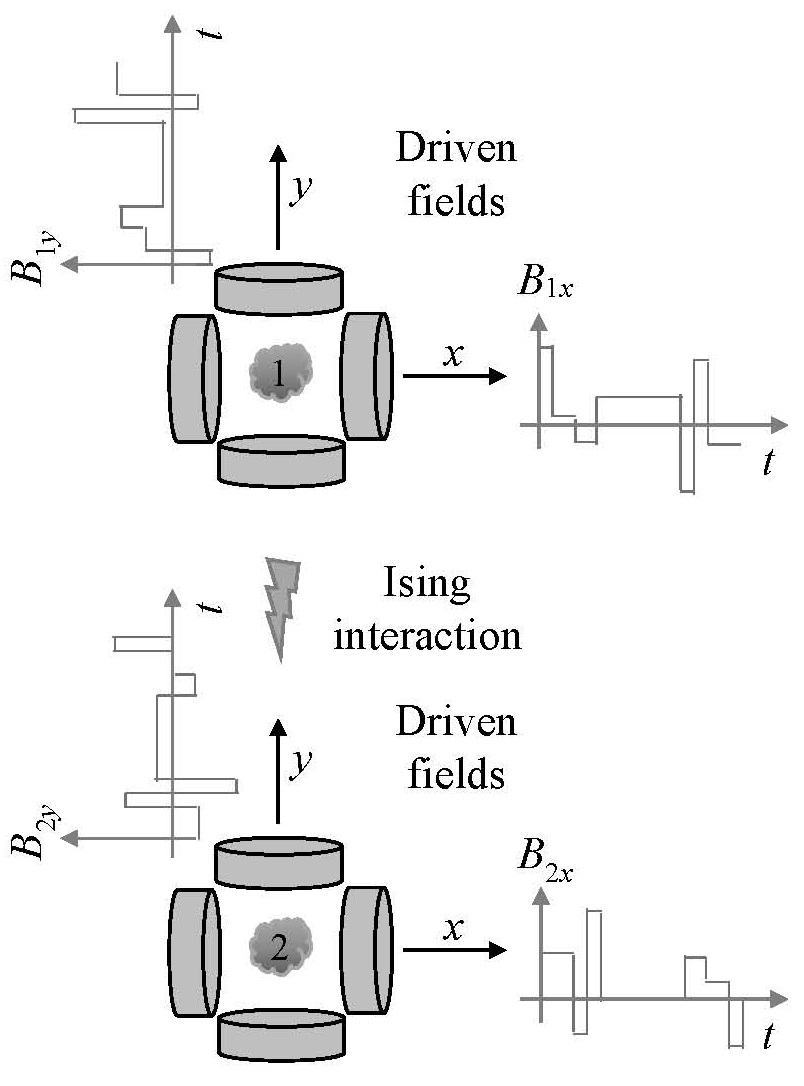}}
\vspace*{8pt}
\caption{Ising interaction mediating between two qubits with local magnetic fields driven the interaction to produce information exchange in the system. When a non local description is used, this basis exhibit a more symmetric form in evolution which can be used as arena for control operations and factorization.}
\label{fig0}
\end{figure}

The use of magnetic systems as quantum resources is a traditional basis on which quantum applications could be settled. Since quantum memories to quantum processors  are considered as stuff for quantum computation or quantum information (so they are a matter susceptible of magnetic control processing). As a simple model of interaction, Ising model brings an easy basis to generate and manipulate quantum states and entanglement particularly. In this model, as is shown in Figure \ref{fig0}, two qubits interact via Ising interaction with additional local magnetic fields as driven elements. 

\subsection{3D anisotropic Ising model and notation}

Recently, some results about anisotropic Ising model generalizes their treatment and suggest an algebraic structure when is depicted on Bell basis \cite{delgadoA}. Following that work, we focus on the following Hamiltonian for the bipartite anisotropic Ising model including an inhomogeneous magnetic field restricted to the $h$-direction ($h=1,2,3$ corresponding with $x,y,z$ respectively):

\begin{eqnarray} \label{hamiltonian}
H_h&=&-\mathbf{\sigma_1 \cdot J \cdot \sigma_2}+\mathbf{B_1 \cdot \sigma_1}+\mathbf{B_2 \cdot \sigma_2} \nonumber \\
&=& -\sum_{k=1}^3 J_k {\sigma_1}_k {\sigma_2}_k+{B_1}_h {\sigma_1}_h+{B_2}_h {\sigma_2}_h\ 
\end{eqnarray}

\noindent which includes several models considered in the before cited works. Following definitions and notation in \cite{delgadoA}: 

\begin{eqnarray} \label{erres}
{R_h}_\pm &=&\sqrt{{B_h^2}_\pm+{J_{i,j}^2}_\mp}=\sqrt{{B_h^2}_\pm+{J_{\{ h \}}^2}_\mp}
\\ \nonumber
{\rm with:} && J_{\{ h \}\pm} \equiv {J_{i,j}}_\pm=J_i \pm J_j \\ \nonumber 
&& B_{h \pm}=B_{1_h}\pm B_{2_h}
\end{eqnarray}

\noindent being $h,i,j$ a cyclic permutation of $1,2,3$ and with pair $i,j$simplified by $\{ h \} \equiv i,j$. As in \cite{delgadoA}, we introduce the scaled parameters:

\begin{eqnarray} \label{defs}
{b_h}_\pm=\frac{{B_h}_\pm}{{R_h}_\pm} &,& {j_h}_\pm=\frac{{J_{\{h\}}}_\mp}{{R_h}_\pm} \in [-1,1]
\end{eqnarray}

\noindent and a nomenclature based on different subscripts than those inherited by the computational basis: greek scripts for $-1,+1$ or $-,+$, for scripts in states and operators (meaning $-1,+1$ in mathematical expressions respectively); capital scripts $A, B, ...$ for $0, 1$ as in the computational basis; and latin scripts $h,i,j,k,...$ for spatial directions $x,y,z$ or $1,2,3$. $\cdot$ will be used as in \cite{delgadoA} for number multiplication in order to avoid misconceptions. Bell states become in this notation:

\begin{eqnarray} \label{bellnotation}
\left| \beta_{--} \right> \equiv \left| \beta_{00} \right> &,& 
\left| \beta_{-+} \right> \equiv \left| \beta_{01} \right> \\ \nonumber 
\left| \beta_{+-} \right> \equiv \left| \beta_{10} \right> &,&  
\left| \beta_{++} \right> \equiv \left| \beta_{11} \right>    
\end{eqnarray}

\subsection{Algebraic structure of evolution operator}

In agreement with last notation, energy levels in (\ref{hamiltonian}) are denoted by eigenvalues $E_{\mu \nu}: E_{--},E_{-+},E_{+-},E_{++}$ respectively:

\begin{eqnarray} \label{eigenvalues2}
{E_h}_{\mu \nu}&\equiv&{\mathcal{E}_h}^{(2+\mu+\frac{1+\nu}{2})}= \mu J_h+\nu {R_h}_{-\mu}  \\ \nonumber
& =&\mu J_h+\nu \sqrt{{B_h}^2_{-\mu}+{J^2_{\{h\}}}_{\mu}}
\end{eqnarray}

\noindent with eigenvectors reported in \cite{delgadoA}. A relevant aspect here is that $U(t) \in SU(4)$ because the sum of eigenvalues is zero. As there, by introducing following definitions:

\begin{equation} \label{delta}
{\Delta_h}_\mu^\nu = \frac{t}{2} ({E_h}_{\mu +}+\nu {E_h}_{\mu -})=
\begin{cases}
\mu J_h t & \rm{if} \quad \nu=+ \\
{R_h}_{-\mu} t & \rm{if} \quad \nu=-
\end{cases}
\end{equation}

\noindent and:

\begin{eqnarray} \label{ed}
{e_h}_\alpha^\beta &=& \cos {\Delta_h}_\alpha^- + i \beta {j_h}_{-\alpha} \sin {\Delta_h}_\alpha^- \\ \nonumber
{d_h}_\alpha &=& {b_h}_{-\alpha} \sin {\Delta_h}_\alpha^-
\end{eqnarray}

Thus, evolution operators in Bell basis are in matrix form:

\begin{widetext}
\begin{eqnarray} \label{mathamiltonian}
{U_{1}}(t)=& \left(
\begin{array}{c|c|c|c}
{e^{i {\Delta_1}_-^+}{e_1}_-^-}^* & i e^{i {\Delta_1}_-^+}{d_1}_-      & 0           & 0      \\
\hline
i e^{i {\Delta_1}_-^+}{d_1}_- & e^{i {\Delta_1}_-^+}{{e_1}_-^-}  & 0           & 0            \\
\hline
0         & 0          & {e^{i {\Delta_1}_+^+}{e_1}_+^+}^*  & -i e^{i {\Delta_1}_+^+}{d_1}_+  \\
\hline
0         & 0          & -i e^{i {\Delta_1}_+^+}{d_1}_+     & {e^{i {\Delta_1}_+^+}{e_1}_+^+} 
\end{array}
\right)  &\in \mathbb{S}^*_1 \\[3mm] \nonumber
{U_{2}}(t)=& \left(
\begin{array}{c|c|c|c}
e^{i {\Delta_2}_+^+}{{e_2}_+^+}^*    &   0   &   0   & - e^{i {\Delta_2}_+^+}{d_2}_+  \\
\hline
0  &  e^{i {\Delta_2}_-^+}{{e_2}_-^+}^* &  -e^{i {\Delta_2}_-^+}{{d_2}_-}  & 0        \\
\hline
0  &  e^{i {\Delta_2}_-^+}{{d_2}_-} &  e^{i {\Delta_2}_-^+}{{e_2}_-^+}  & 0           \\
\hline
e^{i {\Delta_2}_+^+}{d_2}_+    &   0   &   0   & e^{i {\Delta_2}_+^+}{{e_2}_+^+}  
\end{array} 
\right) &\in \mathbb{S}^*_2 \\[3mm] \nonumber
{U_{3}}(t)=& \left(
\begin{array}{c|c|c|c}
e^{i {\Delta_3}_-^+}{{e_3}_-^+}^* & 0 & i e^{i {\Delta_3}_-^+}{d_3}_-      & 0        \\
\hline
0  &  e^{i {\Delta_3}_+^+}{{e_3}_+^+}^* & 0 & i e^{i {\Delta_3}_+^+}{d_3}_+           \\
\hline
i e^{i {\Delta_3}_-^+}{d_3}_- & 0 &  e^{i {\Delta_3}_-^+}{{e_3}_-^+}      & 0         \\
\hline
0  &  i e^{i {\Delta_3}_+^+}{d_3}_+ & 0 & e^{i {\Delta_3}_+^+}{{e_3}_+^+} 
\end{array}
\right) &\in \mathbb{S}^*_3
\end{eqnarray}
\end{widetext}

\noindent with $U_{h}(t)$ have a $2 \times 2$ sector structure and belonging to subgroups of $SU(4)$ (defined in \cite{delgadoA, delgadoB}): $\mathbb{S}^*_1, \mathbb{S}^*_2, \mathbb{S}^*_3$ characterized by some entries equal to zero and having the elements generated by their respective $U_h(t)$ in (\ref{mathamiltonian}). Thus, identity and inverses are included in each subgroup, while there are closure in the product. This group structure is essential in the current work because it assure the existence of solutions for factorization (here, we are simplified the notation for ${{{\mathbb S}^*_h}\substack { \{ |{{j_h}_{\mp \alpha}}| \} \\ \{ s_{{b_h}_{\mp \alpha}} \} }}$ as was used in \cite{delgadoA, delgadoB} by $\mathbb{S}^*_h$ only).

While, sectors are elements of $U(2)$ and as is reported in \cite{delgadoA,delgadoB}, their general structure is:

\begin{equation}\label{sector}
{s_h}_{j} = {e^{i {\Delta_h}_\alpha^+} \left(
\begin{array}{cc}
{{e_h}_\alpha^\beta}^* & -q i^h {d_h}_\alpha   \\
q {i^*}^h {d_h}_\alpha & {{e_h}_\alpha^\beta}    
\end{array}
\right) \vline}_{\tiny \begin{aligned} \alpha &= (-1)^{h+j+1} \\ \beta &= (-1)^{j(h+l_j-k_j+1)} \\ q &= \beta (-1)^{h+1} \end{aligned}}
\end{equation}

\noindent being $h$ their magnetic field direction; $j=1, 2$ the ordering label for sector as it appears in the rows of the evolution matrix: $k_j, l_j$, the labels for its rows (by example, in ${s_2}_1$, $k_2=2, l_2=3$ labels the rows of second sector, $j=2$, in $U_{h=2}(t)$). Note that $\det ({s_h}_{j})={e^{2 i {\Delta_h}_\alpha^+}}$ is unitary.

As we will see, last structure lets introduce the generation of operations in terms of factorization of special unitary matrices in $SU(4)$. As was reported in \cite{delgadoB}, conditions to diagonalize at time $t$ last sector into the form ${s_h}_{j}={\mathbb I}_2$ ($2 \times 2$ identity matrix) is:

\begin{eqnarray} \label{evloop1}
t&=&\frac{2m_\alpha+n_\alpha}{\alpha J_h}\pi >0 \\ \nonumber
{B_h}_{-\alpha}^2&=&(\frac{J_h n_\alpha}{2m_\alpha-n_\alpha})^2-{{J_{\{h\}}}_\alpha}^2 \\ \nonumber
\\ \nonumber
{\rm with:} && n_\alpha,m_\alpha \in \mathbb{Z} \\ \nonumber
\end{eqnarray}

This conditions should be compatible with other restrictions in order to construct a factorization based on $P-$unitary matrices as was developed in \cite{li1}, a sophisticated adaptation of Gauss-Jordan factorization \cite{burden1}.

\section{$P$-unitary matrices factorization}

\subsection{$P-$unitary matrices}

A two level $n-$dimensional $P-$unitary matrix is a unitary matrix obtained departing from $n \times n$ identity matrix, ${\mathbb I}_n$, but including a substitution of some of its elements as follows. If $P=\{j_1,j_2,...,j_n\}$ is a permutation from $\{1,2,...,n\}$, the $P-$unitary matrix $M^n_{j_k,j_{k+1}}$ is said of type $k\in \{1,2,...,n-1\}$ if their entries $(j_k,j_k),(j_k,j_{k+1}),(j_{k+1},j_k),(j_{k+1},j_{k+1})$ were substituted by entries of an arbitrary $2 \times 2$ unitary matrix. In particular, in this work we will be interested in $4 \times 4$ $P-$unitary matrices:

\begin{eqnarray} \label{pmat4}
\nonumber
{M^4_{1,2}}= \left(
\begin{array}{c c c c}
* & *  & 0  & 0      \\
* & *  & 0  & 0      \\
0 & 0  & 1  & 0      \\
0 & 0  & 0  & 1      
\end{array}
\right),   
{M^4_{3,4}}= \left(
\begin{array}{c c c c}
1 & 0  & 0  & 0      \\
0 & 1  & 0  & 0      \\
0 & 0  & *  & *      \\
0 & 0  & *  & *      
\end{array}
\right)  \in {\mathbb S}_1 \\[3mm] \nonumber
{M^4_{1,4}}= \left(
\begin{array}{c c c c}
* & 0  & 0  & *      \\
0 & 1  & 0  & 0      \\
0 & 0  & 1  & 0      \\
* & 0  & 0  & *      
\end{array}
\right),
{M^4_{2,3}}= \left(
\begin{array}{c c c c}
1 & 0  & 0  & 0      \\
0 & *  & *  & 0      \\
0 & *  & *  & 0      \\
0 & 0  & 0  & 1      
\end{array}
\right)  \in {\mathbb S}_2 \\[3mm] \nonumber
{M^4_{1,3}}= \left(
\begin{array}{c c c c}
* & 0  & *  & 0      \\
0 & 1  & 0  & 0      \\
* & 0  & *  & 0      \\
0 & 0  & 0  & 1      
\end{array}
\right),
{M^4_{2,4}}= \left(
\begin{array}{c c c c}
1 & 0  & 0  & 0      \\
0 & *  & 0  & *      \\
0 & 0  & 1  & 0      \\
0 & *  & 0  & *      
\end{array}
\right)  \in {\mathbb S}_3 \\[3mm] 
\end{eqnarray}

\noindent where $*$  means elements substituted in such way that $M^n_{j_k,j_{k+1}} \in U(n)$ are still unitary. It defines a $2 \times 2$ unitary sector embed in the $4 \times 4$ matrix. These matrices have the property that any special unitary matrix $U$ can be written as a product of at most $\frac{n(n-1)}{2}$ $P-$unitary matrices whose determinants product is $\det(U)$. Actually, procedure depicted in \cite{li1} admits $U$ as unitary; we have restricted our analysis to $SU(n)$ because our interest on $U \in SU(4)$ as evolution matrix being generated by (\ref{hamiltonian}).

\subsection{$P-$unitary matrices factorization}

Nevertheless that we focus our analysis to $4 \times 4$  matrices representing interactions in pairs of qubits, we will state the factorization matrix depicted in \cite{li1} for general cases in the following way. We begin with the $n \times n$ unitary matrix, $U$:

\begin{eqnarray}\label{matrixU}
{U}=& \left(
\begin{array}{c c c c c}
a_{1,1} & a_{1,2}  & \hdots  & a_{1,n-1} & a_{1,n}      \\
a_{2,1} & a_{2,2}  & \hdots  & a_{2,n-1} & a_{2,n}     \\
\vdots & \vdots   & \vdots  & \vdots  & \vdots    \\
a_{n-1,1} & a_{n-1,2}  & \hdots  & a_{n-1,n-1} & a_{n-1,n}      \\
a_{n,1} & a_{n,2}  & \hdots  & a_{n,n-1} & a_{n,n}      
\end{array}
\right)  & \\[3mm] \nonumber
\end{eqnarray}

Then, we desire to convert this matrix into ${\mathbb I}_n$ by multiplying it with a series of $P-$unitary matrices. We begin trying to eliminate the $a_{n,1}$ element with one of those $P-$unitary matrices:

\begin{eqnarray}\label{matrixU1}
U^{(n-1,1)} & \equiv & 
\left(
\begin{array}{c c c c c}
1 & 0  & \hdots  & 0 & 0     \\
0 & 1  & \hdots  & 0 & 0     \\
\vdots & \vdots  & \vdots  & 0 & 0     \\
0 & 0  & \hdots & *  & *      \\
0 & 0  & \hdots & *  & *  
\end{array}
\right) \cdot U \\
&=& \left(
\begin{array}{c c c c c}
a_{1,1} & a_{1,2}  & \hdots  & a_{1,n-1} & a_{1,n}      \\
a_{2,1} & a_{2,2}  & \hdots  & a_{2,n-1} & a_{2,n}     \\
\vdots & \vdots   & \vdots  & \vdots  & \vdots    \\
u_{n-1,1} & a^{(n,1)}_{n-1,2}  & \hdots  & a^{(n,1)}_{n-1,n-1} & a^{(n,1)}_{n-1,n}      \\
0 & a^{(n,1)}_{n,2}  & \hdots  & a^{(n,1)}_{n,n-1} & a^{(n,1)}_{n,n}      
\end{array}
\right)  \nonumber\\[3mm] \nonumber
\end{eqnarray}

\noindent where $a^{(i,j)}_{k,l}$ is the transformed entry $k,l$ of matrix when element $i$ in column $j$ is being eliminated. In this sense, superscript is just a reference about the step in which each entry is. Note that in this process, each time that one entry is set to zero, only two rows become modified: the current row which entry is and the immediately before. To reach last outcome, we need that sector in $P-$unitary matrix have the form:

\begin{eqnarray}\label{sectorP1}
\left(
\begin{array}{c c}
\frac{a^*_{n-1,1}}{u_{n-1,1}} & \frac{a^*_{n,1}}{u_{n-1,1}}      \\
-\mu_{n,1} \frac{a_{n,1}}{u_{n-1,1}} & \mu_{n,1} \frac{a_{n-1,1}}{u_{n-1,1}}     
\end{array}
\right)  \\[3mm] \nonumber
\end{eqnarray}

\noindent where $\mu_{n,1}$ is a unitary complex number and:

\begin{eqnarray}\label{sqr1}
u_{n-1,1}=\sqrt{|a_{n-1,1}|^2+|a_{n,1}|^2}
\end{eqnarray}

Continuing this procedure to eliminate upper elements in column $1$, by example to eliminate $a_{n-2,1}$ element:

\begin{eqnarray}\label{matrixU2}
U^{(n-2,1)} & \equiv & \left(
\begin{array}{c c c c c}
1 & \hdots  & 0 & 0 & 0     \\
\vdots & \vdots  & \vdots  & \vdots & \vdots     \\
0 & \hdots  & * & *  & 0      \\
0 & \hdots  & * & *  & 0      \\
0 & \hdots  & 0 & 0  & 1  
\end{array}
\right) \cdot U^{(n-1,1)} \\
&=& \left(
\begin{array}{c c c c c}
a_{1,1} & a_{1,2}  & \hdots  & a_{1,n-1} & a_{1,n}      \\
\vdots & \vdots   & \vdots  & \vdots  & \vdots    \\
u_{n-2,1}   & a^{(n-1,1)}_{n-2,2}  & \hdots  & a^{(n-1,1)}_{n-2,n-1} & a^{(n-1,1)}_{n-2,n}      \\
0 & a^{(n-1,1)}_{n-1,2}  & \hdots  & a^{(n-1,1)}_{n-1,n-1} & a^{(n-1,1)}_{n-1,n}      \\
0 & a^{(n,1)}_{n,2}  & \hdots  & a^{(n,1)}_{n,n-1} & a^{(n,1)}_{n,n}      
\end{array}
\right) \nonumber\\[3mm] \nonumber
\end{eqnarray}

\noindent with sector in $P-$unitary matrix as:

\begin{eqnarray}\label{sectorP2}
\left(
\begin{array}{c c}
\frac{a^*_{n-2,1}}{u_{n-2,1}} & \frac{u^*_{n-1,1}}{u_{n-2,1}}      \\
-\mu_{n-1,1} \frac{u_{n-1,1}}{u_{n-2,1}} & \mu_{n-1,1} \frac{a_{n-2,1}}{u_{n-2,1}}     
\end{array}
\right) \\[3mm] \nonumber
\end{eqnarray}

\noindent where $\mu_{n-1,1}$ is again a unitary complex number and:

\begin{eqnarray}\label{sqr2}
u_{n-2,1}=\sqrt{|a_{n-2,1}|^2+u_{n-1,1}^2}
\end{eqnarray}

Generalizing last procedure to eliminate the $a_{i,j}$ element for $1 \le j < n, j < i \le n$, and denoting by $M^{(i,j)}$ to its corresponding $P-$unitary matrix, we state the following procedure:

\begin{eqnarray}\label{matrixUgral}
U^{(i-1,j)} & \equiv & \left(
\begin{array}{c c c c c c}
1 & \hdots  & 0 & 0 & \hdots & 0     \\
\vdots & \vdots  & \vdots & \vdots  & \vdots & \vdots     \\
0 & \hdots  & * & *  & \hdots &  0     \\
0 & \hdots  & * & *  & \hdots &  0     \\
\vdots & \vdots  & \vdots & \vdots  & \vdots & \vdots     \\
0 & \hdots  & 0 & 0 & \hdots & 1  
\end{array}
\right) \cdot U^{(i,j)} \\
& \equiv & M^{(i,j)} \cdot U^{(i,j)} \nonumber \\
&=& \left(
\begin{array}{c c c c c c}
1 & \hdots  & 0 & 0  & \hdots & 0      \\
\vdots & \vdots &  \vdots & \vdots  & \vdots  & \vdots    \\
0 & \hdots & u_{i-1,j}  &  a^{(i,j)}_{i-1,i-1}  & \hdots & a^{(i,j)}_{n-2,n}      \\
0 & \hdots  & 0  & a^{(i,j)}_{i,j+1} & \hdots & a^{(i,j)}_{i,n}   \\  
\vdots & \vdots &  \vdots & \vdots  & \vdots  & \vdots    \\
0 & \hdots  & 0  & a^{(n,j)}_{n,j+1} & \hdots & a^{(n,j)}_{n,n}      
\end{array}
\right) 
\nonumber\\[3mm] \nonumber
\end{eqnarray}

\noindent as before, we need that sector $s_{(i,j)}$ in $P-$unitary matrix $M^{(i,j)}$ becomes:

\begin{eqnarray}\label{sectorPgral}
s_{(i,j)} &\equiv& \left(
\begin{array}{c c}
\frac{a^{(i-1,j-1)*}_{i-1,j}}{u_{i-1,j}} & \frac{u_{i,j}^*}{u_{i-1,j}}      \\
-\mu_{i,j} \frac{u_{i,j}}{u_{i-1,j}} & \mu_{i,j} \frac{a^{(i-1,j-1)}_{i-1,j}}{u_{i-1,j}}     
\end{array}
\right)  \\[3mm] \nonumber
\end{eqnarray}

\noindent here $a^{(i,0)}_{k,l}=a_{k,l}$ and $\mu_{i,j}$ is again a unitary complex number and:

\begin{eqnarray}\label{sqrgral}
u_{i-1,j}&=&\sqrt{|a^{(j-1,i-2)}_{i-1,j}|^2+u_{i,j}^2} \\
u_{n,j}&=&a_{n,j}^{(n,j-1)}
\end{eqnarray}

With this, we get:

\begin{eqnarray}\label{factorization}
U^{(n-1,n-1)}=\left( \prod^{\leftarrow}_{\substack {1 \le j < n \\ n > i \ge j}} M^{(i,j)} \right)  U
\end{eqnarray}

\noindent where symbol $\leftarrow$ states that product is the backward product which stacks factors from right to left according to order of scripts. For the case that we are interested, $U$ is unitary and each $M^{(i,j)}$ are unitary too because $\det (M^{(i,j)})=\mu_{i,j}$ is unitary and finally we get $U^{(n-1,n-1)}$ unitary. But it implies that their rows and columns are unitary. It means, when first column of entries below of diagonal are set to zero, automatically row $1$ becomes $1$ on the diagonal and zero off diagonal because of $u_{1,1}=1$. But clearly, as each $U^{(j,j)}$ is unitary too, then its block off diagonal in columns and rows are eliminated until this step. Last property repeats for each column eliminated in spite of $u_{j,j}$ definition:

\begin{eqnarray}\label{u}
u_{j,j}&=&\sqrt{\sum_{j \le i \le n}|a^{(i,j-1)}_{i,j}|^2}
\end{eqnarray}

\noindent then, with that:

\begin{eqnarray}\label{det}
\det (U^{(n-1,n-1)})=\prod_{\substack {1 \le j < n \\ j<i\le n}} \mu_{i,j} det(U)=1
\end{eqnarray}
 
The way to fulfill (\ref{det}) is open in principle, but it can be used to fit specific requirements in the construction of each $M^{(i,j)}$. Finally, the main result of this section is that:

\begin{eqnarray}\label{muproduct}
U=\prod^{\rightarrow}_{\substack {1 \le j < n \\ n > i \ge j}} M^{(i,j)\dagger} 
\end{eqnarray}

\noindent where symbol $\rightarrow$ states that product is the forward product which stacks factors from left to right. It procedure means that some unitary gate or interaction could be reproduced by a series of $P-$unitary matrix ${M^{(i,j)}}^\dagger$ if it is possible give the adequate form to each one. Clearly, when any entry will be zero, then process skip their elimination, which means that respective matrix ${M^{(i,j)}}$ is ${\mathbb I}_4$.

\section{Solutions based on Ising model evolution}

\subsection{Tuning Ising interaction to $P-$unitary matrices based on one pulse} \label{iva}

For $4 \times 4$ order for matrix evolution, as was written in last section, $M^4_{j_k,j_{k+1}} \in {\mathbb{S}^*_i}$, it means that at least, they have the form of Ising evolution matrices when they are written in Bell basis. In particular in the process depicted, only $U_1(t)$ and $U_2(t)$ are useful to reduce them in those forms, which have the structure of matrices ${M^{(i,j)}}^\dagger$. Now, the challenge is to analyze if sector (\ref{sector}) could be fitted to (\ref{sectorPgral}) and their subsidiary requirements.

By reviewing sectors to fit, (\ref{sector}) and (\ref{sectorPgral}), comparing entries in ${{s_h}_\alpha}$ with entries in ${s_{(i,j)}}^\dagger$ we get the following set of equations for each entry:

\begin{eqnarray}\label{secequations1p}
1,1: &&  {{e_h}_\alpha^\beta}^*{e^{i {\Delta_h}_\alpha^+}}=\frac{a^{(i-1,j-1)}_{i-1,j}}{u_{i-1,j}} \nonumber \\
1,2: &&  -q i^h {d_h}_\alpha {e^{i {\Delta_h}_\alpha^+}}=-\mu_{i,j}^* \frac{u_{i,j}^*}{u_{i-1,j}}   \nonumber \\
2,1: &&  q {i^*}^h {d_h}_\alpha {e^{i {\Delta_h}_\alpha^+}}= \frac{u_{i,j}}{u_{i-1,j}}  \nonumber \\
2,2: &&  {{e_h}_\alpha^\beta} {e^{i {\Delta_h}_\alpha^+}}=\mu_{i,j}^* \frac{{a^{(i-1,j-1)}_{i-1,j}}^*}{u_{i-1,j}} \nonumber \\
\end{eqnarray}

\noindent so we find a faithful concordance. The first requirement appears by comparing equations of sectors $1,1$ with $2,2$, and $1,2$ with $2,1$ too:

\begin{eqnarray}\label{eqmu}
\mu_{i,j}={e^{-2 i {\Delta_h}_\alpha^+}}
\end{eqnarray}

\noindent with that, only equations for sectors $1,1$ and $2,1$ should be solved. Yet, because of (\ref{det}), in agreement with forms in (\ref{mathamiltonian}) and the process previously depicted, the following global restriction should be fulfilled:

\begin{eqnarray}\label{global}
{\Delta_x}_{+,(4,1)}^+ + {\Delta_y}_{-,(3,1)}^+ + {\Delta_x}_{-,(2,1)}^+ + &\quad& \nonumber \\
\quad  {\Delta_x}_{+,(4,2)}^+ + {\Delta_y}_{-,(3,2)}^+ + {\Delta_x}_{+,(4,3)}^+ &=& N \pi \nonumber \\
\end{eqnarray}

\noindent with $N \in {\mathbb Z}$. Here, the subscripts $(i,j)$ are related with the entry being eliminated. In spite of first equation in (\ref{evloop1}) should be fulfilled in order to fit the remaining sector to the identity (by changing $\alpha$ by $-\alpha$ in (\ref{evloop1})), then ${\Delta_h}_{\alpha,(i,j)}^+$ becomes an integer multiple of $\pi$, so (\ref{global}) is trivially satisfied. Nevertheless of last general explanation, clearly since beginning we can note that $U_h(t)$ in (\ref{mathamiltonian}) is in $SU(4)$, then obviously $\mu_{i,j}=1$ and condition (\ref{global}) is fulfilled easier. As a consequence, this procedure based on Ising interactions only can generate $U \in SU(4)$, being consistently with them.

If $a^{(i-1,j-1)}_{i-1,j} \equiv |a^{(i-1,j-1)}_{i-1,j}|e^{i {\phi^{(i-1,j-1)}_{i-1,j}}}$ and $u_{i,j} \equiv |u_{i,j}|e^{i {\varphi_{i,j}}}$ (note that a phase different from zero applies only for $u_{n,j}$, the first element to be eliminated in the bottom of each column), then by splitting the restrictions for phases and magnitudes we find the next three equivalent conditions:

\begin{eqnarray}
|\chi|&\equiv&\frac{|u_{i,j}|}{u_{i-1,j}}=|{b_h}_{-\alpha} \sin {\Delta_h}_\alpha^-| \label{secequations1pb} \\ 
{\phi^{(i-1,j-1)}_{i-1,j}}&=&{\Delta_h}_\alpha^+ - \tan^{-1} ({j_h}_{-\alpha} \beta \tan {\Delta_h}_\alpha^-) \label{secequations1pc} \\
{\varphi_{i,j}}&=&{\Delta_h}_\alpha^+ - \frac{\pi}{2} (h-1+{\rm sign} (q \chi)) \label{secequations1pd}
\end{eqnarray}

This procedure implies to solve these equations for the non diagonal sector  combining them with identity sector requirements in each $P-$unitary matrix of (\ref{muproduct}) in order to fit them to adequate forms (\ref{pmat4}). Thus, each matrix becomes an evolution boost obtained by means of magnetic field pulse of specific duration in the Ising hamiltonian (\ref{hamiltonian}). Unfortunately, this procedure fails to give solutions in spite of (\ref{secequations1pd}), which can not be fulfilled. Because ${\Delta_h}_\alpha^+=-{\Delta_h}_{-\alpha}^+$ and the remaining sector should be reduced to ${\mathbb I_2}$, in agreement with \cite{delgadoB}, ${\Delta_h}_\alpha^+=-(2m_{-\alpha}+n_{-\alpha})\pi$, with $m_{-\alpha}, n_{-\alpha} \in {\mathbb Z}$, being ${\Delta_h}_{-\alpha}^- = n_{-\alpha} \pi$. Then for $h=1$ it is impossible that $\varphi_{i,j}=0$ or the specific value required by $\varphi_{n,j}$. 

For $h=2$, equation (\ref{secequations1pd}) is easily fulfilled for matrices $U^{(i,j)}$ with real entries by selecting adequately $n_{-\alpha}$: $\varphi_{i,j}=(n_{-\alpha}-\frac{1}{2}(1+{\rm sign}(q \chi) )\pi$. By defining $\theta \equiv {{J_{\{h\}}}_\alpha} t$ and $|\chi| \equiv {|u_{i,j}|}/{u_{i-1,j}} \in [0,1]$, (\ref{secequations1pb}) can be written as: 

\begin{eqnarray}\label{secequations1pe}
\chi&=& {\rm sign}({b_h}_{-\alpha}) \sqrt{1-\frac{\theta^2}{{{\Delta^2_h}_\alpha^-}}} \sin {\Delta_h}_\alpha^-
\end{eqnarray}

\begin{figure}[pb] 
\centerline{\psfig{width=3.3in,file=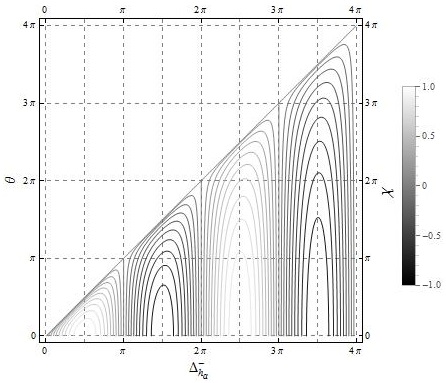}}
\vspace*{8pt}
\caption{Solution curves of (\ref{secequations1pb}) for different values of $\chi \in [-1,1]$ indicated by gray levels as shows the $\chi$-scale on right. They are limited to the down diagonal half in the first quadrant.} 
\label{fig1}
\end{figure}

Some solutions are shown in Figure \ref{fig1} for values of $\chi \in [-1,1]$ (note that $0 \le \chi \le 1$ because $|u_{i,j}| \le u_{i-1,j}$) which is reported as a gray level in the color of graphs. Range of values for ${\Delta_h}_\alpha^-$ becomes limited while $|\chi|$ increases. Because of (\ref{secequations1pe}),  values of ${\Delta_h}_\alpha^-$ when $\theta=0$ determines the range by continuous sections of ${\Delta_h}_\alpha^-$ variable. These ranges are fixed for each value of $\chi$ independently of intervals $[(n-1) \pi,n \pi],n \in {\mathbb Z}^+$. Thus, when $|\chi|$ increases, it is more difficult that (\ref{secequations1pc}) and (\ref{secequations1pd}) be fulfilled in general because range of ${\Delta_h}_\alpha^-$ shrinks. In addition, as for real entries case, equation (\ref{secequations1pc}) requires that $\tan{\Delta_h}_\alpha^- =0$ or ${j_h}_{-\alpha}=0$ (then $|{b_h}_{-\alpha}|=1$). First restriction is impossible except for $|\chi|=0$ (which is not useful) and second one is only applicable for partially anisotropic or isotropic interactions (${J_{\{y\}}}_{\alpha}=J_z-J_x=0$). In this last case, $|\chi|=|\sin {\Delta_h}_\alpha^-|$, so matrices $M^{(3,1)}$ and $M^{(3,2)}$ can be constructed with this particular case for Ising interaction with driven magnetic fields in $y-$direction. Additional prescriptions are given by (\ref{evloop1}), by changing $\alpha$ with $-\alpha=+1$ in agreement with (\ref{mathamiltonian}) for diagonal sector in $U_2(t)$ to fit it to $M^4_{2,3}$. For non diagonal sector with $\alpha=-1$, it is required that ${\rm sign}(a^{(i-1,j-1)}_{i-1,j})=(-1)^{n_{-\alpha}}{\rm sign}(\cos{\Delta_h}_\alpha^-)$, ${\rm sign}(u_{i,j})=(-1)^{n_{-\alpha}-\frac{1}{2}(1+{\rm sign}(q \chi))}=(-1)^{n_{-\alpha}-\frac{1}{2}(1-{\rm sign}(\chi))}$ (which could be a possible strong restriction because $\varphi_{i,j}$) and:

\begin{eqnarray}\label{prescription1p}
{B_h}_-=\frac{1}{t}\sin^{-1}\frac{u_{i,j}}{u_{i-1,j}}
\end{eqnarray}

\noindent where election of inverse for sine function should be selected in order to fulfill previous relations for signs of $a^{(i-1,j-1)}_{i-1,j},u_{i,j}$. Summarizing, one pulse solutions are in the most cases unable to generate a set of Ising evolutions which become factors of a general unitary operation. Thus, more complex schemes should be analyzed.

\subsection{Tuning Ising interaction to $P-$unitary matrices based on two pulses} \label{ivb}

As was shown in \cite{delgadoA}, evolutions given by (\ref{mathamiltonian}) form a Lie group and in addition, two combined pulses let to get any $U(2)$ element for each sector (more precisely, two ${s_h}_j$ consecutively sectors can reproduce any $SU(2)$ element already considering (\ref{global})). Here, one sector ($-\alpha$) should be driven to ${\mathbb I}_2$ while remaining sector ($\alpha$) reproduces the $P-$unitary matrix as part of factorization presented in \cite{li1}. Thus, is warranted that two pulses should to generate matrices (\ref{pmat4}).

\subsubsection{Equations to get $P-$unitary matrix factors in two pulses}

When we combine two pulses, the generic form of combined sector is:

\begin{widetext}
\begin{eqnarray} \label{twopulses}
{s'_h}_{j} {s_h}_{j}={e^{i ({\Delta'_h}_\alpha^+ + {\Delta_h}_\alpha^+)}} \times  
 \left(
\begin{array}{cc}
{{e'_h}_\alpha^\beta}^*{{e_h}_\alpha^\beta}^* - {d'_h}_\alpha {d_h}_\alpha & 
-q i^h ({{e'_h}_\alpha^\beta}^*{d_h}_\alpha + {{e_h}_\alpha^\beta}{d'_h}_\alpha )   \\
q {i^*}^h ({{e'_h}_\alpha^\beta}{d_h}_\alpha + {{e_h}_\alpha^\beta}^*{d'_h}_\alpha ) & 
{{e'_h}_\alpha^\beta}{{e_h}_\alpha^\beta} - {d'_h}_\alpha {d_h}_\alpha    
\end{array}
\right)
\end{eqnarray}
\end{widetext}

\noindent as is reported in \cite{delgadoA,delgadoB}. Repeating the process as in the one pulse case, the only relevant equations for non-diagonal sector $\alpha$ desired in $U'_h(t')U_h(t)$ are those for the entries:

\begin{eqnarray}\label{secequations2p}
1,1: &&  ({{e'_h}_\alpha^\beta}^*{{e_h}_\alpha^\beta}^* - {d'_h}_\alpha {d_h}_\alpha) e^{i ({\Delta'_h}_\alpha^+ + {\Delta_h}_\alpha^+)}=\frac{a^{(i-1,j-1)}_{i-1,j}}{u_{i-1,j}} \nonumber \label{secequations2pa} \\
2,1: &&  q {i^*}^h ({{e'_h}_\alpha^\beta}{d_h}_\alpha + {{e_h}_\alpha^\beta}^*{d'_h}_\alpha ) e^{i ({\Delta'_h}_\alpha^+ + {\Delta_h}_\alpha^+)}=\frac{u_{i,j}}{u_{i-1,j}}   \nonumber \label{secequations2pb} \\
\end{eqnarray}
 
\noindent and as before, it is required that:

\begin{eqnarray} \label{eqmu2p}
\mu_{i,j}=e^{-2i ({\Delta'_h}_\alpha^+ + {\Delta_h}_\alpha^+)}
\end{eqnarray}

\noindent in order to fulfill automatically those equations for $1,2$ and $2,2$ entries. As before, we can divide equations (\ref{secequations2pa}) in four equations, two for their magnitude and two for the phases. Because of each sector (\ref{twopulses}) is unitary, then magnitude equations will be equivalent, so we take only the $2,1$ entry magnitude equation. With that, we have the following three equations for the non-diagonal sector $\alpha$:

\begin{eqnarray}
|\chi_h| &\equiv& \frac{|u_{i,j}|}{u_{i-1,j}} = |{{e'_h}_\alpha^\beta}{d_h}_\alpha + {{e_h}_\alpha^\beta}^*{d'_h}_\alpha| \label{secequations2p12m} \\ 
{\varphi_{i,j}}&=&{\varphi'_{i,j}} + ({\Delta'_h}_\alpha^+ + {\Delta_h}_\alpha^+) - \nonumber \\
&& \frac{\pi}{2} (h-1+{\rm sign} (q)) \label{secequations2p12f} \\
{\phi^{(i-1,j-1)}_{i-1,j}}&=&{\phi'^{(i-1,j-1)}_{i-1,j}}+({\Delta'_h}_\alpha^+ + {\Delta_h}_\alpha^+)  \label{secequations2p11f}
\end{eqnarray}

\noindent where ${\phi'^{(i-1,j-1)}_{i-1,j}}=\arg({{e'_h}_\alpha^\beta}^*{{e_h}_\alpha^\beta}^* - {d'_h}_\alpha {d_h}_\alpha)$ and ${\varphi'_{i,j}}=\arg({{e'_h}_\alpha^\beta}{d_h}_\alpha + {{e_h}_\alpha^\beta}^*{d'_h}_\alpha)$. Note that last phase make a difference with respect to one pulse impossibility. This phase is generated as consequence of ${s_h}_j \in U(2)$ but does not generate $U(2)$, it means, not all elements in such group can be obtained from ${s_h}_j$ \cite{delgadoA}, then a product of two of them, ${s'_h}_j, {s_h}_j$, are in $U(2)$ but is not a third ${s''_h}_j$, having a different structure in $U(2)$. Last equations should be solved together with prescriptions to get ${\mathbb I}_2$ for the remaining sector with label $-\alpha$. In agreement with \cite{delgadoB}, prescriptions to get ${\mathbb I}_2$ in the sector $-\alpha$ in a matrix evolution for two pulses are:

\begin{eqnarray}\label{diag}
&{{\Delta_h}_{-\alpha}^-} +{\rm sign}({J_{\{h\}}}_{-\alpha}{J'_{\{h\}}}_{-\alpha}) {\Delta'_h}_{-\alpha}^-=n_{-\alpha} \pi \nonumber \\ \nonumber
&{\Delta_h}_{-\alpha}^+ + {\Delta'_h}_{-\alpha}^+= (2m_{-\alpha}+n_{-\alpha}) \pi \\ 
&\frac{{B'_h}_{\alpha}}{{J'_{\{h\}}}_{-\alpha}}=\frac{{B_h}_{\alpha}}{{J_{\{h\}}}_{-\alpha}} \\ \nonumber
{\rm with:} & m_{-\alpha}, n_{-\alpha} \in \mathbb{Z} \nonumber \\ \nonumber
\end{eqnarray}

Note that equation (\ref{global}) is still required. Because ${\Delta_h}_{-\alpha}^+ + {\Delta'_h}_{-\alpha}^+=-({\Delta_h}_{\alpha}^+ + {\Delta'_h}_{\alpha}^+)$ and second last equation, (\ref{global}) is again fulfilled automatically. The process to fit each matrix ${M^{(i,j)}}^\dagger$ by means of a two pulse driven Ising interaction is use equations (\ref{diag}) to get $t', {B_h}_{-\alpha},{B'_h}_{-\alpha}$ in terms of $t$. All these equations are easily solved. In particular:

\begin{eqnarray}\label{tp}
t'&=&-\frac{J_h}{J'_h}t-\frac{\alpha (2m_{-\alpha}+n_{-\alpha}) \pi}{J'_h} \nonumber \\
 {{B_h}_\alpha}  &=& \pm \sqrt{ \left(\frac{n_{-\alpha}\pi}{ t + S_\alpha t'}\right)^2-{{J^2_{\{h\}}}_{-\alpha}} } \nonumber \\
{\rm where:} & S_\alpha & = \frac{{J'_{\{h\}}}_{-\alpha}}{{J_{\{h\}}}_{-\alpha}}
\end{eqnarray}

After, equations (\ref{secequations2pa}) should be solved together to get $t, {B_h}_{-\alpha},{B'_h}_{-\alpha}$. Relevant equations equivalent to (\ref{secequations2p12m}-\ref{secequations2p11f}) can be expressed in terms of ${j_h}_{-\alpha},{j'_h}_{- \alpha},{b_h}_{- \alpha},{b'_h}_{- \alpha}, {\Delta_h}_{\alpha}^-,{\Delta'_h}_{\alpha}^-$. In particular main expressions are the real and the imaginary parts in ${{e'_h}_\alpha^\beta}{d_h}_\alpha + {{e_h}_\alpha^\beta}^*{d'_h}_\alpha$:

\begin{eqnarray}\label{expressionsmag}
{\mathcal R}_\varphi &\equiv&   {b_h}_{-\alpha} \sin {\Delta_h}_{\alpha}^- \cos {\Delta'_h}_{\alpha}^- + {b'_h}_{-\alpha} \cos {\Delta_h}_{\alpha}^- \sin {\Delta'_h}_{\alpha}^- \nonumber \\
{\mathcal I}_\varphi &\equiv& \beta \sin {\Delta_h}_{\alpha}^- \sin {\Delta'_h}_{\alpha}^- ({j'_h}_{-\alpha}{b_h}_{-\alpha}-{j_h}_{-\alpha}{b'_h}_{-\alpha}) \nonumber \\
\end{eqnarray}

\noindent and for ${{e'_h}_\alpha^\beta}^*{{e_h}_\alpha^\beta}^* - {d'_h}_\alpha {d_h}_\alpha$:

\begin{eqnarray}
{\mathcal R}_\phi &\equiv& \cos {\Delta_h}_{\alpha}^- \cos {\Delta'_h}_{\alpha}^- - \nonumber \\
&& \quad ({j'_h}_{-\alpha} {j_h}_{-\alpha} + {b'_h}_{-\alpha} {b_h}_{-\alpha}) \sin {\Delta_h}_{\alpha}^- \sin {\Delta'_h}_{\alpha}^- \nonumber \\
{\mathcal I}_\phi &\equiv& - \beta ({j'_h}_{-\alpha} \sin {\Delta'_h}_{\alpha}^- \cos {\Delta_h}_{\alpha}^- + \nonumber \\
&& \quad {j_h}_{-\alpha} \sin {\Delta_h}_{\alpha}^- \cos {\Delta'_h}_{\alpha}^-)
\end{eqnarray}

Noting that ${\mathcal R}_\varphi^2+{\mathcal I}_\varphi^2+{\mathcal R}_\phi^2+{\mathcal I}_\phi^2=1$. The procedure will require to solve (\ref{secequations2p12m}) in the form:

\begin{eqnarray}\label{chi}
{\chi_h}^2 = {{\mathcal R}_\varphi}^2+{{\mathcal I}_\varphi}^2 &=& \frac{|u_{i,j}|^2}{u_{i-1,j}^2}  \label{ec1tosolvea} 
\end{eqnarray}

\noindent together with equations (\ref{secequations2p12f}-\ref{secequations2p11f}), which can be rewritten as:

\begin{eqnarray}\label{varphi}
{\varphi_{i,j}} &=& \arctan \xi_\varphi + \frac{\pi}{2} (1- {\rm sign} ({\mathcal R}_\varphi)) -\\ 
&& (2 m_{-\alpha}+n_{-\alpha}) \pi + \frac{\pi}{2} (h-1+{\rm sign} (q)) \nonumber \\
{\phi^{(i-1,j-1)}_{i-1,j}} &=& \arctan \xi_\phi + \frac{\pi}{2}(1-{\rm sign} ({\mathcal R}_\phi)) - \\
&& (2 m_{-\alpha}+n_{-\alpha}) \pi \nonumber \\
&{\rm with:}& \xi_\varphi \equiv \frac{{\mathcal I}_\varphi}{{\mathcal R}_\varphi}, \xi_\phi \equiv \frac{{\mathcal I}_\phi}{{\mathcal R}_\phi} \nonumber
\end{eqnarray}

Last equations shows that if ${\chi_h}^2$ covers $[0,1]$ independently of $\xi_\varphi, \xi_\phi$ values, then (\ref{sector}) can be adapted always to (\ref{sectorPgral}) with two pulses. Note particularly that sign in ${\mathcal R}_\varphi, {\mathcal R}_\phi$ can be easily adapted, so last requirements are centered on $\xi_\varphi, \xi_\phi$, which should to have, each one, free ranges in ${\mathbb R}$ independently to ${\chi_h}$. Additionally, these equations depends on relative signs between ${j_h}_{-\alpha},{j'_h}_{-\alpha},{b_h}_{-\alpha},{b'_h}_{-\alpha}$: $s_{j_h} \equiv {\rm sign}({j_h}_{-\alpha} {j'_h}_{-\alpha}), s_{b_h} \equiv {\rm sign}({b_h}_{-\alpha} {b'_h}_{-\alpha})$. Resultant equations:

\begin{widetext}
\begin{eqnarray}
\xi_\varphi &=& \frac{\beta \sin {\Delta_h}_{\alpha}^- \sin {\Delta'_h}_{\alpha}^- ({j'_h}_{-\alpha}{b_h}_{-\alpha}-{j_h}_{-\alpha}{b'_h}_{-\alpha})}{{b_h}_{-\alpha} \sin {\Delta_h}_{\alpha}^- \cos {\Delta'_h}_{\alpha}^- + {b'_h}_{-\alpha} \cos {\Delta_h}_{\alpha}^- \sin {\Delta'_h}_{\alpha}^-} \label{eqstunning1} \\
\xi_\phi &=& \frac{- \beta ({j'_h}_{-\alpha} \sin {\Delta'_h}_{\alpha}^- \cos {\Delta_h}_{\alpha}^- + 
 {j_h}_{-\alpha} \sin {\Delta_h}_{\alpha}^- \cos {\Delta'_h}_{\alpha}^-)}{\cos {\Delta_h}_{\alpha}^- \cos {\Delta'_h}_{\alpha}^- -  ({j'_h}_{-\alpha} {j_h}_{-\alpha} + {b'_h}_{-\alpha} {b_h}_{-\alpha}) \sin {\Delta_h}_{\alpha}^- \sin {\Delta'_h}_{\alpha}^-} \label{eqstunning2} \\
{\chi_h}^2 &=& (1+\xi_\varphi^2)({b_h}_{-\alpha} \sin {\Delta_h}_{\alpha}^- \cos {\Delta'_h}_{\alpha}^- + {b'_h}_{-\alpha} \cos {\Delta_h}_{\alpha}^- \sin {\Delta'_h}_{\alpha}^-)^2 \label{eqstunning3} 
\end{eqnarray}
\end{widetext}

\noindent involves several parameters, but states the general procedure to obtain matrix factors. Thus, it should be probed that those equations can be fulfilled by any $\chi_h, \varphi_{i,j}$ and ${\phi^{(i-1,j-1)}_{i-1,j}}$, together with (\ref{diag}-\ref{tp}), to set the precise values for ${b_h}_{\pm \alpha},{b'_h}_{\pm \alpha}, {\Delta_h}_{\alpha}^-,{\Delta'_h}_{\alpha}^-$ (or their equivalent parameters ${B_h}_{\pm \alpha},{B'_h}_{\pm \alpha},t,t'$). 

\subsubsection{Integrating requirements to obtain solutions for $P-$unitary matrix factors}

Note that previous set of equations become in a non-linear equations system in spite of ${\Delta_h}_{\alpha}^-, {\Delta'_h}_{\alpha}^-$ contains ${B_h}_{- \alpha},{B'_h}_{- \alpha}$. Parameters should be uncoupled because in spite of (\ref{diag}-\ref{tp}), there are only three free parameters remaining to solve three equations (\ref{chi}-\ref{varphi}): $t,{B_h}_{-\alpha},{B'_h}_{-\alpha}$. Thus, equation for $t'$ in (\ref{secequations2p11f}) should be explicitly used in (\ref{chi}-\ref{varphi}) and ${\Delta_h}_{\alpha}^-, {\Delta'_h}_{\alpha}^-$ written conveniently:

\begin{eqnarray}\label{teqs}
{\Delta_h}_{\alpha}^- &=& \frac {{R_h}_{-\alpha}} {{J_{\{h\}}}_{\alpha}} {J_{\{h\}}}_{\alpha} t \equiv \frac{\tau}{{j_h}_{-\alpha}} \nonumber \\
{\Delta'_h}_{\alpha}^- &=& \frac {{R'_h}_{-\alpha}} {{J'_{\{h\}}}_{\alpha}} {{J'_{\{h\}}}_{\alpha}} t' \equiv \frac{\tau'}{{j'_h}_{-\alpha}} \nonumber \\
\tau' &=& - \tau \frac {{J'_{\{h\}}}_{\alpha}}{J'_h} \frac {J_h} {{J_{\{h\}}}_{\alpha}}  - \alpha (2m_{-\alpha}+n_{-\alpha}) \pi \frac {{J'_{\{h\}}}_{\alpha}}{J'_h} \nonumber \\
&\equiv& - \tau \frac{c'_\alpha}{c_{\alpha}}+c'_\alpha N_\alpha \pi
\end{eqnarray}

\noindent where $c_\alpha$ denotes the ratio between transverse strength ${J_h}_{\{\alpha\}}$ and parallel strength ${J_h}$. $N_{-\alpha} = -\alpha(2m_{-\alpha}+n_{-\alpha}) \in {\mathbb Z}$. Parameters ${j_h}_{-\alpha} \in [-1,1],{j'_h}_{-\alpha}\in [-1,1], \tau \in {\mathbb R}$ appears together with $s_{j_h},s_{b_h},c_\alpha, c'_\alpha, N_{-\alpha}$ constants in the resultant equations (observe that signs in ${j_h}_{-\alpha},{j'_h}_{-\alpha}$ could not be selected, instead, they depend on quantum system; nevertheless it is not a problem because properties of trigonometric functions which warranted multiple solutions with signs changed in $\xi_\varphi, \xi_\phi$). Non-linear nature of equations require a numerical treatment to find general solutions. Still, each set of required values, $\chi_h,\xi_\varphi, \xi_\phi$ to reproduce a specific $P-$unitary factor matrix of $U \in SU(4)$, is expected to have multiple solutions. 

\subsubsection{Existence of solutions for $P-$unitary matrices factorization in two pulses}

\begin{figure}[pb] 
\centerline{\psfig{width=2.2in,file=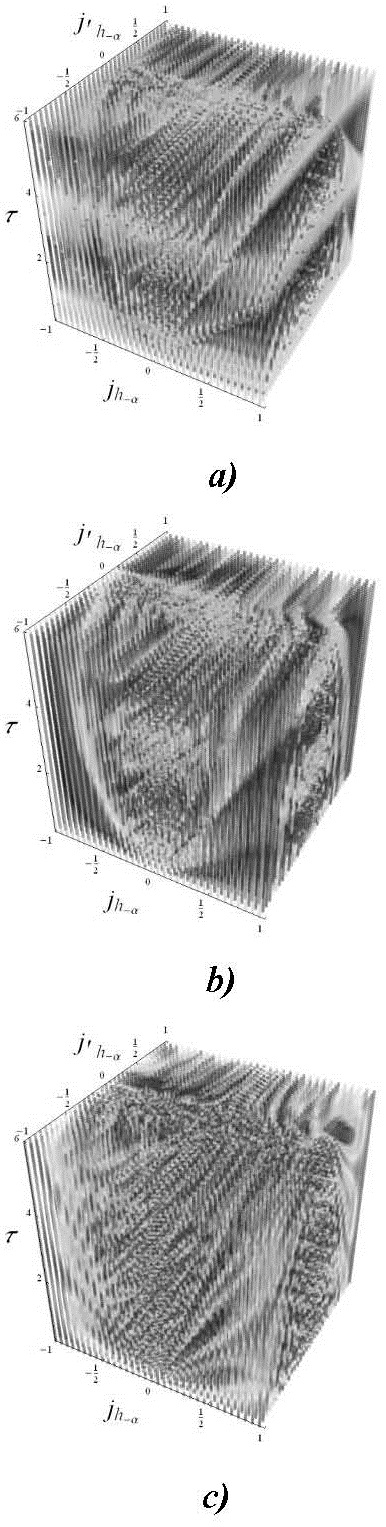}}
\vspace*{8pt}
\caption{Colored maps of a) $\arctan(\xi_\varphi) \in (-\pi/2,\pi/2)$, b) $\arctan(\xi_\phi) \in (-\pi/2,\pi/2)$, and c) $\chi_h^2 \in [0,1]$. Dark colors correspond to lower values in their respective range and brighter to upper ones. All graphs are represented in $({j_h}_{-\alpha},{j'_h}_{-\alpha},\tau) \in [-1,1]\times[-1,1]\times[0,6]$. Maps correspond to case $s_{j_h}=s_{b_h}=1,c_\alpha=c'_\alpha=1, N_{-\alpha}=2$, as an example.}
\label{fig2}
\end{figure}

Clearly numerator and denominator in expressions (\ref{eqstunning1}-\ref{eqstunning2}), in terms of (\ref{teqs}), have still a lot of possibilities to null independently and non simultaneously, so because continuity, $\xi_\varphi,\xi_\phi$ ranges in ${\mathbb R}$. In addition, $\chi_h^2$ clearly ranges in $[0,1]$. Here, we take care about apparent correlation between null denominator in $\xi_\varphi$ and $\chi_h$, this is only due to written form of $\chi_h^2$ in (\ref{eqstunning3}). Unfortunately, equations (\ref{eqstunning1}-\ref{eqstunning3}) are a family, depending on physical parameters $s_{j_h},s_{b_h},c_\alpha, c'_\alpha, N_{-\alpha}$ by considering (\ref{teqs}). In addition, these parameters can not be integrated in only one non dimensional parameter, leaving a unique set of general equations. Figure \ref{fig2} shows colored maps representing $\arctan(\xi_\varphi),\arctan(\xi_\phi), \chi_h^2$ in terms of ${j_h}_{-\alpha},{j'_h}_{-\alpha},\tau$ for $s_{j_h}=s_{b_h}=1,c_\alpha=c'_\alpha=1, N_{-\alpha}=2$. In them, dark colors represent lower values and bright colors to higher values in their respective ranges ($(-\pi/2,\pi/2)$ for $\arctan(\xi_\varphi),\arctan(\xi_\phi)$ and $[0,1]$ for $\chi_h^2$). These maps show the non-linear complexity of those equations. 

Nevertheless complexity, their solution is warranted because of connectivity of sector elements ${s_h}_j$ in (\ref{mathamiltonian}) through their finite products as part of Lie group $U(2)=U(1) \times SU(2)$, which form the groups ${\mathbb S}^*_h$ when those sectors are adequately combined, as is discussed in \cite{delgadoA, delgadoB}. Still, some of that solutions could correspond in some cases to non physical or experimentally complex situations, as an example $|{b_h}_{\pm \alpha}|,|{b'_h}_{\pm \alpha}|=1$. Because there are lot of direct and indirect parameters involved, the best strategy is generate a computational procedure to solve this problem. First, a procedure which states the form of each $P$-unitary matrix factor,  ${M^{(i,j)}}^\dagger$, and then apply on each one a numerical procedure to solve equations (\ref{diag},\ref{tp},\ref{eqstunning1}-\ref{teqs}) together, in which, parameters as $s_{j_h}, s_{b_h},c_\alpha, c'_\alpha, N_{-\alpha}$ and others should be selected. Because there are multiple roots, a stochastic procedure to find an specific solution into the correspondent graphs as those in Figure \ref{fig2} is more practical. Then, this procedure lets to find solutions to problem of factorization of unitary evolution matrices $U$ with complex entries based on Ising interaction pulses, being restricted by consistence to $det(U)=1$.

\section{Special case: unitary matrices with real entries}

A special, but very common case, could be attended: special unitary matrices $U$ whose entries are real. Fortunately, this case exhibit an analytical easier solution which is depicted in this section in terms of theory previously developed until here. 

\subsection{Analytical solution for $P-$unitary matrix factors with $h=1$}

To obtain ${M^{(4,1)}}^\dagger, {M^{(2,1)}}^\dagger, {M^{(4,2)}}^\dagger$ and ${M^{(4,3)}}^\dagger$, we have as restriction, the impossibility depicted in section \ref{iva} for one pulse: to have all entries as real because factors ${i^*}^h,{i^h}$ in antidiagonal entries. An alternative is try to get all entries as imaginary, which is possible by select ${{\Delta_h}_\alpha^-}=\frac{2n_\alpha+1}{2} \pi$. With this selection, we get for non diagonal sector:

\begin{eqnarray}\label{sectorh1}
{s_h}_{j} &=& -i e^{i {\Delta_h}_\alpha^+} (-1)^{n_\alpha}  \left(
\begin{array}{cc}
\beta {j_h}_{-\alpha} & q {b_h}_{-\alpha}  \\
q {b_h}_{-\alpha} & -\beta {j_h}_{-\alpha}    
\end{array} 
\right) \nonumber \\
\end{eqnarray}

 Unfortunately, the impossibility remains because still, $i$ factor can not be eliminated by selecting $e^{i {\Delta_h}_\alpha^+}$ as imaginary, because it requires be real in order to remaining diagonal sector becomes ${\mathbb I}_2$. Thus, analysis with two pulses shows that solutions are possible. Effectively, we can analyze this problem in terms of general procedure in section \ref{ivb} stating ${\mathcal R}_\varphi={\mathcal I}_\phi=0$ with which we find several analytic solutions. One of them is useful here, but in this case, that solution is easier achievable by taking two pulses ${s_h}_j,{s'_h}_j$ with form (\ref{sectorh1}), to obtain:

\begin{eqnarray}\label{sectorh1twop}
{s_h}_{j} &=& e^{i ({\Delta'_h}_\alpha^+)+{\Delta_h}_\alpha^+)} (-1)^{n'_\alpha+n_\alpha+1}  \left(
\begin{array}{cc}
A_+ & -q \beta A_-   \\
q \beta  A_- & A_+ 
\end{array} 
\right) \nonumber \\
{\rm with:}&&A_+ ={j_h}_{-\alpha} {j'_h}_{-\alpha}+{b_h}_{-\alpha}{b'_h}_{-\alpha} \nonumber \\
&&A_-={j_h}_{-\alpha}{b'_h}_{-\alpha}-{b_h}_{-\alpha}{j'_h}_{-\alpha} \nonumber \\
&&A_+^2+A_-^2 = 1
\end{eqnarray}

\noindent in which the uncomfortable factor $i$ clearly disappears. Here, ${{\Delta_h}_\alpha^-}=\frac{2n_\alpha+1}{2}\pi \equiv \frac{N_\alpha}{2} \pi,{{\Delta'_h}_\alpha^-}=\frac{2n'_\alpha+1}{2}\pi \equiv \frac{N'_\alpha}{2} \pi$. Still, the following equations should be solved in order to get ${b_h}_{-\alpha},{b'_h}_{-\alpha}$:

\begin{eqnarray}\label{solh1}
m^{(i,j)}_{1,1} &\equiv& \frac{a^{(i-1,j-1)}_{i-1,j}}{u_{i-1,j}} = (-1)^S  ({j_h}_{-\alpha} {j'_h}_{-\alpha}+{b_h}_{-\alpha}{b'_h}_{-\alpha})
\nonumber \\
m^{(i,j)}_{2,1} &\equiv& \frac{u_{i,j}}{u_{i-1,j}} = \beta q (-1)^S ({j_h}_{-\alpha}{b'_h}_{-\alpha}-{b_h}_{-\alpha}{j'_h}_{-\alpha})\nonumber \\
\end{eqnarray}

\noindent where $S \equiv {2m_{-\alpha}+n_{-\alpha}+n_{\alpha}+n'_{\alpha}+1}$ retrieve all integer constants which appears in the procedure. Terms $m^{(i,j)}_{1,1},m^{(i,j)}_{2,1}$, for short, correspond to entries in each matrix ${M^{(i,j)}}^\dagger$. It is clear that those equations can be reduced to first one by noting that ${m^{(i,j)}_{1,1}}^2+{m^{(i,j)}_{2,1}}^2=1$. At end, both equations just require a review about concordance in signs. Parameters in $S$ can be selected to obtain this concordance. Still, it is required to write $t,t'$ in terms of $\frac{2n_\alpha+1}{2},\frac{2n'_\alpha+1}{2}$ and ${R_h}_{\alpha},{R'_h}_{\alpha}$. Then, by substituting them in first equation in (\ref{tp}):

\begin{eqnarray}\label{eq}
{j_h}_{-\alpha}\frac{N_\alpha}{2c_\alpha}+{j'_h}_{-\alpha}\frac{N'_\alpha}{2c'_\alpha}&=&-{\alpha (2m_{-\alpha}+n_{-\alpha})} \nonumber \\
\end{eqnarray}

\noindent all in terms of ${j_h}_{-\alpha}, {j'_h}_{-\alpha}$. As before, nevertheless that we will not introduce time explicitly here, it is convenient note that definitions introduced in (\ref{teqs}) for non dimensional time $\tau={j_h}_{-\alpha}\frac{N_\alpha}{2}\pi,\tau'={j'_h}_{-\alpha}\frac{N'_\alpha}{2}\pi$ are prevailing to report specific results in terms of non dimensional parameters ${j_h}_{\pm\alpha}, {j'_h}_{\pm\alpha},{b_h}_{\pm\alpha}, {b'_h}_{\pm\alpha},\tau,\tau'$ without involve physical parameters ${J_{\{h\}}}_{\pm \alpha},{J'_{\{h\}}}_{\pm \alpha},{J_h},{J'_h}$. At this point, note that for diagonal sector:

\begin{eqnarray}\label{solh2jm}
|{j_h}_\alpha| &=& |{j'_h}_\alpha|=\frac{1}{n_{-\alpha}\pi}(\tau \frac{c_{-\alpha}}{c_{\alpha}}+\tau' \frac{c'_{-\alpha}}{c'_{\alpha}}) \nonumber \\
|{b_h}_\alpha| &=& |{b'_h}_\alpha|
\end{eqnarray}

\noindent where similarly to $c_{\alpha},c'_{\alpha}$, we define $c_{-\alpha}=\frac{{J_{\{h\}}}_{-\alpha}}{J_h},c'_{-\alpha}=\frac{{J'_{\{h\}}}_{-\alpha}}{J'_h}$. Signs in ${j_h}_\alpha, {j'_h}_\alpha$ are physical prescriptions and there are not restrictions in signs of ${b_h}_\alpha, {b'_h}_\alpha$. In further applications, we assume these formulas for two pulse case in order to obtain diagonal sector complementary prescriptions, which will not be explicitly reported because they need ${c_{\pm\alpha}},{c'_{\pm\alpha}}$ specification.  

Thus, both equations (\ref{solh1}) and (\ref{eq}) require be solved simultaneously for ${j_h}_\alpha$ and ${j'_h}_\alpha$. This problem reduce to solve a quadratic equation whose solutions can be expressed as:

\begin{eqnarray}\label{solh2j}
{j_h}_{-\alpha} &=& \frac {C(A+D) \pm |B| \sqrt{B^2-C^2+(A+D)^2}}{B^2+(A+D)^2} \nonumber \\
{j'_h}_{-\alpha} &=& \frac {E(A+F) \pm |B| \sqrt{B^2-E^2+(A+F)^2}}{B^2+(A+F)^2}\nonumber \\
{\rm with:} && A=(-1)^S {m^{(i,j)}_{1,1}}, B={m^{(i,j)}_{2,1}} \nonumber \\
&& C=\frac{2N_{-\alpha}c'_\alpha}{N'_{\alpha}}, E=\frac{2N_{-\alpha}c_\alpha}{N_{\alpha}}\nonumber \\ && D=F^{-1}=\frac{N_{\alpha}c'_\alpha}{N'_{\alpha}c_\alpha}
\end{eqnarray}

\noindent where solutions should be selected by reviewing correct signs in both equations (\ref{solh1}) together with election of signs in ${b_h}_{-\alpha},{b'_h}_{-\alpha}$. Last means that there are not necessarily correspondence between signs in both formulas (\ref{solh2j}).  At end is possible to find all parameters involved to generate ${M^{(i,j)}}^\dagger$ by means of $U_2(t)$. Formula (\ref{solh2jm}) should be analyzed more carefully, because $|{j_h}_\alpha|, |{j'_h}_\alpha| \le 1$, then a detailed view to current set of formulas shows that while existence of solutions in (\ref{solh2j}) depends on greater values for ${N_\alpha},{N'_\alpha}$. But at time, $|{j_h}_\alpha|, |{j'_h}_\alpha|$ values increase with this. Thus, some control could be required to have lower values of $c_{-\alpha},c'_{-\alpha}$.

\subsection{Analytical solution for $P-$unitary matrix factors with $h=2$}

As was stated before, this case admits an easy solution in just one pulse by selecting ${{j_h}_{-\alpha}}=0, |{{b_h}_{-\alpha}}|=1$ (requiring strictly control on ${J_{\{h\}}}_\alpha$ value). Independently of possible experimental restrictions to manipulate ${J_{\{h\}}}_\alpha$ values, this selection gives a practical solution for non diagonal sector:

\begin{eqnarray}\label{sectorh2}
{s_h}_{j} &=& (-1)^{n_{-\alpha}} \cdot \nonumber \\
&& \left(
\begin{array}{cc}
\cos {B_h}_{-\alpha} t & q {\rm sign} ({b_h}_\alpha) \sin {B_h}_{-\alpha} t  \\
-q {\rm sign} ({b_h}_\alpha) \sin {B_h}_{-\alpha} t & \cos {B_h}_{-\alpha} t    
\end{array} 
\right) \nonumber \\
\end{eqnarray}

\noindent in terms of parameters of section \ref{iva}. It implies to obtain $t$ and ${B_h}_{\alpha}$ from (\ref{evloop1}) in order to obtain ${\mathbb I}_2$ in remaining diagonal sector and then solve:

\begin{eqnarray}\label{solh2}
{B_h}_{-\alpha}&=&\frac{1}{t} \cos^{-1}  \frac{(-1)^{n_{-\alpha}} a^{(i-1,j-1)}_{i-1,j}}{u_{i-1,j}} \nonumber \\
{\rm with:} \quad \quad \quad \quad && \nonumber \\
{\rm sign}({B_h}_{-\alpha}) &=& (-1)^{n_{-\alpha}+1} q {\rm sign} ( \frac{u_{i,j}}{u_{i-1,j}} \sin {B_h}_{-\alpha} t ) \nonumber \\
\end{eqnarray}

\noindent to obtain ${B_h}_{-\alpha}$. This equation is equivalent to equation (\ref{prescription1p}). This procedure works to obtain ${M^{(3,1)}}^\dagger, {M^{(3,2)}}^\dagger$. No more solutions in two pulses are able because they require the same condition ${{j_h}_{-\alpha}}=0$ in order to have the form (\ref{sectorPgral}). In order to report results in terms of non dimensional parameters and because here ${{j_h}_{-\alpha}}=0$, terms as $\tau$ are not appropriate. Thus, we introduce the variables $\tau_0 \equiv J_h t$ as in (\ref{evloop1}) with $\alpha \rightarrow -\alpha$, and $b_0 \equiv \frac{{B_h}_{-\alpha}}{J_h}$ (the equivalent expression for (\ref{solh2}) in these terms is immediate) to report results for sector $\alpha$ in which is located the non diagonal sector. For diagonal identity sector, $-\alpha$, the use of ${j_h}_\alpha$ is still appropriate and it can be written as ${j_h}_\alpha=\frac{N_{-\alpha} c_{-\alpha}}{n_{-\alpha}}$, where as for two pulses case: $N_{-\alpha}=-\alpha(2m_{-\alpha}+n_{-\alpha}), c_{-\alpha}=\frac{{J_{\{h\}}}_{-\alpha}}{J_h}$. As is usual, $|{b_h}_\alpha|=\sqrt{1-{j_h}_\alpha^2}$ and its sign can be selected arbitrarily. We will assume these formulas for one pulse case to obtain and report complementary diagonal sector prescriptions (with exception of ${b_h}_\alpha$). 

\subsection{Existence of solutions for $P-$unitary matrices factorization for unitary $U$ with real entries in two pulses}

Formulas for $h=2$ are easy and their only issue is the restriction ${j_h}_{-\alpha}=0$ which implies some control in the interaction strength of interaction (\ref{hamiltonian}). In addition, the correct election of signs depicted in formula (\ref{solh2}) is a trivial issue because of properties of trigonometric functions. 

For $h=1$, lots of parameters appear as in the general case for complex entries, but as we are shown, analytical solutions are possible. As values of $c_\alpha,c'_\alpha$, signs of ${j_h}_{-\alpha},{j'_h}_{-\alpha}$ are not eligible because they are physical constants of system. Still, by example, if $|B|>|C|,C>0, A+D>0$ then ${j_h}_{-\alpha}>0$; instead, if $|B|>|C|,C>0, A+D<0$ then ${j_h}_{-\alpha}<0$. Similar conditions let the same for ${j_h}_{-\alpha}$. Then, a brief analysis shows that this expressions let obtain both signs for ${j_h}_{-\alpha},{j'_h}_{-\alpha}$ by an adequate selection of $N_{-\alpha} \in {\mathbb Z}$ and $N_{\alpha}, N'_{\alpha}$ both odd. Normally, it is possible note that changing signs in $N_\alpha,N'_\alpha$ implies a sign change in ${j_h}_{-\alpha},{j'_h}_{-\alpha}$ solutions. In addition, solutions are physically meaningful ($|{j_h}_{-\alpha}|,|{j'_h}_{-\alpha}|\le1$) if we select higher values for $|N_{\alpha}|, |N'_{\alpha}|$ because it reduces $C,E$ values. By these reasons, normally large values in $N_{-\alpha}$ do not give physical solutions. As was mentioned before, last election could give $|{j_h}_{-\alpha}|,|{j'_h}_{-\alpha}|>1$ because $D$ and $F$ values increases. This could be avoided if $c_\alpha,c_\alpha$ can be manipulated.

In the following section a couple of examples will be developed to shown how the procedure of factorization presented can be implemented.

\section{Applications and examples}

Next examples show applications of previous development by factorizing some special unitary matrices in $SU(4)$ which are not achievable for exclusive interactions in only one group ${\mathbb S}^*_h$, instead by a combination of elements of three groups. Results in \cite{delgadoA} and this work establish that a general operation $U \in SU(4)$ can be reproduced by a finite product of elements in ${\mathbb S}^*_1, {\mathbb S}^*_2$ Ising interactions via factorization proposed in \cite{li1}.

In the following analysis we will assume that factorization is being developed by a system with some physical properties as: a) interaction strength constants positive (or zero if is required), ${j_h}_{\pm \alpha},{j'_h}_{\pm \alpha}$, b) relative strength ratios were set as $c_{\alpha},c'_{\alpha}=1,c_{\alpha},c'_{\alpha}=0.5$. This setting is just to report concrete results more than real restrictions, which, as was discussed, are not present (with exceptions of ${j_h}_{- \alpha}=0$ to get factors for $h=2$ case). Because multiplicity of solutions in terms of several parameters involved $m_{-\alpha},n_{- \alpha},n_{\alpha},n'_{\alpha}$ (or some of their associated values as $N_{\pm \alpha}, N'_{\alpha}$), normally we are selected the lower ones possible. They are included in the prescriptions. Note finally, in particular, that because of definitions of $\tau, \tau'$, they can become negative (no in present case with ${j_h}_{\pm \alpha},{j'_h}_{\pm \alpha}$ positive), remembering that they are no physical time and positiveness of $t,t'$ is always recovered.

\subsection{Equivalent gate to $C^1NOT_2$ gate in Bell basis}

$C^1NOT_2$ gate is a very common and useful gate used in Quantum Computation. A similar gate with determinant $1$ is the controlled gate $C^1(i Y_2)$. This gate can be reproduced in the present scheme noting that this gate, in Bell basis (\ref{bellnotation}), has the form:

\begin{eqnarray} \label{c1not2}
U &=& \frac{1}{2} \left(
\begin{array}{c c c c}
1 & -1  & 1  & 1      \\
1 & 1  & -1  & 1      \\
1 & 1  & 1  & -1      \\
-1 & 1  & 1  & 1      
\end{array}
\right)  \\ \nonumber \\
&=& ({\mathcal H}_1 \otimes {\mathbb I}_2) C^1 NOT_2 (C^1(i Y_2)) C^1 NOT_2 ({\mathcal H}_1 \otimes {\mathbb I}_2) \nonumber 
\end{eqnarray}

$C^1NOT_2$ is a gate sometimes difficult to reproduce, but $C^1(i Y_2)$ is an alternative for this case, which could be generated with physical processes as in the current work through magnetic pulses. Quantum algorithms using $C^1NOT_2$gates could be traduced in terms of $C^1(i Y_2)$. Following the process depicted in the previous sections, we can decompose it in several parts by means of factorization in terms of a set of $P-$unitary matrices achievable by physical interactions. Thus, their $P-$unitary matrix factors and respective design parameters are shown in Table \ref{tabla1} in those terms presented in last section.

\begin{table}[htb] 
 \centering \caption{Factorization in $P-$unitary matrices for $C^1(i Y_2)$ and design parameters of non diagonal sector.} \label{tabla1}
\begin{tabular}{|c|c|}
    \hline
$P-$unitary matrices & Design parameters  \\
    \hline

$\begin{array}{c}
{M^{(4,1)}}^\dagger \\ 
\left(
\begin{array}{cccc}
 1 & 0 & 0 & 0 \\
 0 & 1 & 0 & 0 \\
 0 & 0 & \frac{1}{\sqrt{2}} & -\frac{1}{\sqrt{2}} \\
 0 & 0 & \frac{1}{\sqrt{2}} & \frac{1}{\sqrt{2}} \\
\end{array}
\right)
\end{array}$ & 

$\begin{array}{l}
N_{-\alpha}=-1 \\
j_\alpha=j'_\alpha=0.681 \\
N_{\alpha}=3, N'_{\alpha}=-3 \\
j_{-\alpha}=0.120,b_{-\alpha}=0.993 \\
j'_{-\alpha}=0.787,b'_{-\alpha}=0.617 \\
\tau=0.568,\tau'=3.709   \\
\end{array}$

     \\
    \hline
		
$\begin{array}{c}
{M^{(3,1)}}^\dagger \\
\left(
\begin{array}{cccc}
 1 & 0 & 0 & 0 \\
 0 & \frac{1}{\sqrt{3}} & -\sqrt{\frac{2}{3}} & 0 \\
 0 & \sqrt{\frac{2}{3}} & \frac{1}{\sqrt{3}} & 0 \\
 0 & 0 & 0 & 1 \\
\end{array}
\right)
\end{array}$ &

$\begin{array}{l}
m_{-\alpha}=-2, n_{-\alpha}=3 \\
j_\alpha=0.167 \\
b_0=0.696 \\
\tau_0=3.142   \\
\end{array}$

		\\
    \hline

$\begin{array}{c}
{M^{(2,1)}}^\dagger \\
\left(
\begin{array}{cccc}
 \frac{1}{2} & -\frac{\sqrt{3}}{2} & 0 & 0 \\
 \frac{\sqrt{3}}{2} & \frac{1}{2} & 0 & 0 \\
 0 & 0 & 1 & 0 \\
 0 & 0 & 0 & 1 \\
\end{array}
\right)
\end{array}$ &

$\begin{array}{l}
N_{-\alpha}=-1 \\
j_\alpha=j'_\alpha=0.968 \\
N_{\alpha}=3, N'_{\alpha}=-3 \\
j_{-\alpha}=0.312,b_{-\alpha}=0.950 \\
j'_{-\alpha}=0.978,b'_{-\alpha}=0.205 \\
\tau=1.471,\tau'=4.613   \\
\end{array}$

	  \\
    \hline

$\begin{array}{c}
{M^{(4,2)}}^\dagger \\
\left(
\begin{array}{cccc}
 1 & 0 & 0 & 0 \\
 0 & 1 & 0 & 0 \\
 0 & 0 & \frac{1}{2} & -\frac{\sqrt{3}}{2} \\
 0 & 0 & \frac{\sqrt{3}}{2} & \frac{1}{2} \\
\end{array}
\right)
\end{array}$ &

$\begin{array}{l}
N_{-\alpha}=-1 \\
j_\alpha=j'_\alpha=0.692 \\
N_{\alpha}=3, N'_{\alpha}=-3 \\
j_{-\alpha}=0.128,b_{-\alpha}=-0.992 \\
j'_{-\alpha}=0.795,b'_{-\alpha}=0.607 \\
\tau=0.603,\tau'=3.745   \\
\end{array}$

	  \\
    \hline
		
$\begin{array}{c}
{M^{(3,2)}}^\dagger \\
\left(
\begin{array}{cccc}
 1 & 0 & 0 & 0 \\
 0 & -\frac{1}{\sqrt{3}} & -\sqrt{\frac{2}{3}} & 0 \\
 0 & \sqrt{\frac{2}{3}} & -\frac{1}{\sqrt{3}} & 0 \\
 0 & 0 & 0 & 1 \\
\end{array}
\right)
\end{array}$ &

$\begin{array}{l}
m_{-\alpha}=-2, n_{-\alpha}=3 \\
j_\alpha=0.167 \\
b_0=0.696 \\
\tau_0=3.142   \\
\end{array}$

    \\
    \hline
		
$\begin{array}{c}
{M^{(4,3)}}^\dagger \\
\left(
\begin{array}{cccc}
 1 & 0 & 0 & 0 \\
 0 & 1 & 0 & 0 \\
 0 & 0 & \frac{1}{\sqrt{2}} & -\frac{1}{\sqrt{2}} \\
 0 & 0 & \frac{1}{\sqrt{2}} & \frac{1}{\sqrt{2}} \\
\end{array}
\right)
\end{array}$ &

$\begin{array}{l}
N_{-\alpha}=-1 \\
j_\alpha=j'_\alpha=0.681 \\
N_{\alpha}=3, N'_{\alpha}=-3 \\
j_{-\alpha}=0.120,b_{-\alpha}=0.993 \\
j'_{-\alpha}=0.787,b'_{-\alpha}=0.617 \\
\tau=0.568,\tau'=3.709   \\
\end{array}$

    \\
    \hline
   \end{tabular}
    \end{table}

In \cite{delgadoA,delgadoB} this gate was used to design control gates or to mimic an alternative teleportation quantum algorithm based in Ising interaction based on group properties which assure a coverage of a finite product of elements in ${\mathbb S}^*{_h}$ groups on a maximal group ${\mathbb S}_h$ (treatment there is a little different because $C^1(i Y_2)$ is constructed directly to operate on a grammar based on Bell states instead on typical computational basis). That, together with current example suggest that specialized gates can be constructed via factorization by using Ising interaction. Still, in terms of group theory, more research about a minimal representation in these finite products in ${\mathbb S}_h$ it is necessary.

\subsection{Characterization operation}

Quantum control of physical states requires to change properties of quantum states. A characterization operation \cite{delgadoC} is an unitary operation which change the superposition state of an initial state. By example, the following matrix represent one of these operations (numbers in it are casual to illustrate the factorization process using Ising boosts):

\begin{eqnarray} \label{charac}
U= \frac{1}{10} \left(
\begin{array}{cccc}
 7 & 1 & 7 & -1 \\
 1 & -7 & 1 & 7 \\
 7 & -1 & -7 & -1 \\
 1 & 7 & -1 & 7 \\
\end{array}
\right)
\end{eqnarray}

This operation based on Bell basis lets to transform some initial Bell state (or a superposition of them) into other state. We can convert several initial states with this operation and if we use it repeatedly, this kind of operations have several behaviors in terms of convergence. Convergence of characterization matrices powers is an interesting issue to explode control of quantum states. Following a similar process to find the design parameters of their $P-$unitary factors, we can reproduce these kind of operations by factorization. For specific case (\ref{charac}), these factors and parameters are reported in table \ref{tabla2}. 

\begin{table}[htb] 
 \centering \caption{Factorization in $P-$unitary matrices for characterization matrix $U$ and design parameters of non diagonal sector.} \label{tabla2}
\begin{tabular}{|c|c|}
    \hline
$P-$unitary matrices & Design parameters  \\ \hline

$\begin{array}{c}
{M^{(4,1)}}^\dagger \\
\left(		
\begin{array}{cccc}
 1 & 0 & 0 & 0 \\
 0 & 1 & 0 & 0 \\
 0 & 0 & \frac{7}{5 \sqrt{2}} & -\frac{1}{5 \sqrt{2}} \\
 0 & 0 & \frac{1}{5 \sqrt{2}} & \frac{7}{5 \sqrt{2}} \\
\end{array}
\right)
\end{array}$ & 

$\begin{array}{l}
N_{-\alpha}=-1 \\
j_\alpha=j'_\alpha=0.500 \\
N_{\alpha}=-3, N'_{\alpha}=-3 \\
j_{-\alpha}=0.400,b_{-\alpha}=0.916 \\
j'_{-\alpha}=0.267,b'_{-\alpha}=0.964 \\
\tau=1.886,\tau'=1.256   \\
\end{array}$

     \\
    \hline

$\begin{array}{c}
{M^{(3,1)}}^\dagger \\ 
\left(
\begin{array}{cccc}
 1 & 0 & 0 & 0 \\
 0 & \frac{1}{\sqrt{51}} & -5 \sqrt{\frac{2}{51}} & 0 \\
 0 & 5 \sqrt{\frac{2}{51}} & \frac{1}{\sqrt{51}} & 0 \\
 0 & 0 & 0 & 1 \\
\end{array}
\right) 
\end{array}$ &

$\begin{array}{l}
m_{-\alpha}=-2, n_{-\alpha}=3 \\
j_\alpha=0.167 \\
b_0=0.548 \\
\tau_0=3.142   \\
\end{array}$

		\\
    \hline

$\begin{array}{c}
{M^{(2,1)}}^\dagger \\ 
\left(
\begin{array}{cccc}
 \frac{7}{10} & -\frac{\sqrt{51}}{10} & 0 & 0 \\
 \frac{\sqrt{51}}{10} & \frac{7}{10} & 0 & 0 \\
 0 & 0 & 1 & 0 \\
 0 & 0 & 0 & 1 \\
\end{array}
\right) 
\end{array}$ &

$\begin{array}{l}
N_{-\alpha}=-1 \\
j_\alpha=j'_\alpha=0.968 \\
N_{\alpha}=-3, N'_{\alpha}=3 \\
j_{-\alpha}=0.978,b_{-\alpha}=0.205 \\
j'_{-\alpha}=0.312,b'_{-\alpha}=0.950 \\
\tau=4.612,\tau'=1.471   \\
\end{array}$

	  \\
    \hline

$\begin{array}{c}
{M^{(4,2)}}^\dagger \\ 
\left(
\begin{array}{cccc}
 1 & 0 & 0 & 0 \\
 0 & 1 & 0 & 0 \\
 0 & 0 & \frac{7}{10} & -\frac{\sqrt{51}}{10} \\
 0 & 0 & \frac{\sqrt{51}}{10} & \frac{7}{10} \\
\end{array}
\right)
\end{array}$ &

$\begin{array}{l}
N_{-\alpha}=-1 \\
j_\alpha=j'_\alpha=0.704 \\
N_{\alpha}=-3, N'_{\alpha}=3 \\
j_{-\alpha}=0.802,b_{-\alpha}=0.596 \\
j'_{-\alpha}=0.136,b'_{-\alpha}=0.991 \\
\tau=3.783,\tau'=0.641   \\
\end{array}$

	  \\
    \hline

$\begin{array}{c}
{M^{(3,2)}}^\dagger \\ 
\left(
\begin{array}{cccc}
 1 & 0 & 0 & 0 \\
 0 & -\frac{1}{\sqrt{51}} & -5 \sqrt{\frac{2}{51}} & 0 \\
 0 & 5 \sqrt{\frac{2}{51}} & -\frac{1}{\sqrt{51}} & 0 \\
 0 & 0 & 0 & 1 \\
\end{array}
\right) 
\end{array}$ &

$\begin{array}{l}
m_{-\alpha}=-2, n_{-\alpha}=3 \\
j_\alpha=0.167 \\
b_0=0.455 \\
\tau_0=3.142   \\
\end{array}$

    \\
    \hline

$\begin{array}{c}
{M^{(4,3)}}^\dagger \\ 
\left(
\begin{array}{cccc}
 1 & 0 & 0 & 0 \\
 0 & 1 & 0 & 0 \\
 0 & 0 & \frac{7}{5 \sqrt{2}} & -\frac{1}{5 \sqrt{2}} \\
 0 & 0 & \frac{1}{5 \sqrt{2}} & \frac{7}{5 \sqrt{2}} \\
\end{array}
\right)
\end{array}$ &

$\begin{array}{l}
N_{-\alpha}=-1 \\
j_\alpha=j'_\alpha=0.500 \\
N_{\alpha}=-3, N'_{\alpha}=-3 \\
j_{-\alpha}=0.400,b_{-\alpha}=0.916 \\
j'_{-\alpha}=0.267,b'_{-\alpha}=0.964 \\
\tau=1.886,\tau'=1.256   \\
\end{array}$

    \\
    \hline
\end{tabular}
\end{table}

This example shows that for a general characterization matrix:

\begin{eqnarray} \label{characgral}
U= \left(
\begin{array}{cccc}
 \alpha & \beta & \gamma & -\delta \\
 \beta & -\alpha & \delta & \gamma \\
 \gamma & -\delta & -\alpha & -\beta \\
 \delta & \gamma & -\beta & \alpha \\
\end{array}
\right)
\end{eqnarray}

\noindent will exist a process of quantum modeling based on factorization. A rich research field based on power bounded and power convergence for unitary matrices is open, in order to control this procedures when they are based on Ising pulses. An additional treatment associated with quantum error correction is recommended for this kind of operations.

\section{Conclusions}

Current research about physical systems on which set up quantum technology, in particular those related with quantum computation, quantum information processing or quantum cryptography, are growing from several directions to get them in a useful form for applications. Nevertheless that optics has been partially a dominant arena to last developments, matter has been shown several own benefits in some involved aspects. 

Quantum storage and quantum information processing allow new computational tasks which are impossible with conventional information technology or quantum optics exclusively. In these trends, systems based on trapped ions, e-Helium, nuclear magnetic resonance, superconductors, doped silicon and quantum dots have shown opportunities to make stable and efficient developments for that purpose \cite{klo1}. Such quantum stuff requires a system of several qubits. The main materials based technology known for that is magnetic. For this reason, spin-based quantum computing has been developed in several experimental implementations, which uses magnetic systems mainly: superconducting integrated systems, superconducting flux qubits, straight nuclear magnetic resonance and quantum dots. All of them exploits Ising interactions with different approaches \cite{john1}, together with control on quantum states and in particular with entanglement control, a milestone in all almost these researches.

Circuit-gate model was the first approach to quantum computation, nevertheless, quantum annealing \cite{kado1} or measurement-based quantum computation \cite{briegel1} are alternatives which use magnetic systems approached by Ising interactions to manage a planned and controlled quantum state manipulation. On them, several applied problems has been exhibited as the goal of these technologies (pattern matching, folding proteins, an other particular NP-complete problems \cite{john1}) in order to test them.

In spite of current work, for magnetic systems and under the regular perspective of interaction when is depicted on entangled pairs basis, the direct change of classical computational basis on an entangled basis as their basic computer grammar should be considered, inclusively with actual control problems around decoherence entanglement. Thus, another extension is clearly an analogous analysis for multiqubit systems in terms of an adequate basis of entangled states (as $\left| GHZ \right>$, $\left| W \right>$ and other entangled states in $SU(8)$ or still greater systems) as in the model presented here. In this work, solutions (\ref{solh2jm}-\ref{solh2}) set a concrete theoretical method to generate gates with real entries based on Ising pulses for two qubits computation. More generally, procedure (\ref{eqstunning1}-\ref{teqs}) states a numerical method to solve same problem for complex entries gates in general. As is stated in \cite{delgadoA}, an important extension of results presented there for multiqubit interactions, it is obtain their evolution in terms of sectors relating their 'natural' states. Last will let extensions of factorization stated in \cite{li1} in terms of Ising interaction pulses as was implemented in this work, being extended for a higher number of qubits when they need be processed simultaneously.

In the same way, error correction analysis is necessary in our procedures, based on error factors in control (magnetic field and time measurement, precise knowledge about interaction strengths, etc.). In this line of research, the analysis of behavior with finite temperature based on matrix density is mandatory to consider decoherence effects. Finally, it is required improvement through alternative continuous pulses to generate the same effects and factorization procedures as those presented here. Nevertheless that magnetic rectangular pulses are easy to manage theoretically and still experimentally too because currently there are a tight control on them (reducing  their   resonant effects because their discontinuity), the best is seek continuous fields (\cite{delgadoB} has proposed $B(t)=B_0+B_p \sin \omega t$) to try a reproduction of this kind of effects in (\ref{hamiltonian}). This method, if works, could suppose to apply directed waves on the matter in order to generate in it certain quantum operations based on factorization.  

Nuclear magnetic resonance, Quantum dots and Electrons in silicon lattices are named as the most successful systems in implementing quantum algorithms based on their coherence and stability. Experimental applications to set up current proposal on these technologies should develop a narrow orbit control and strength interactions control. Currently, this capacities of control are still far but they are emerging. Then, a more deep control could be applied in terms of present work to generate entanglement control and induced gates design on matter. 


\small  

\end{document}